\documentclass[11pt,a4paper]{article}
\usepackage[hyperref]{eacl2021}
\usepackage{times}
\usepackage{latexsym}

\usepackage{microtype}

\usepackage{multirow,booktabs,amsmath,url,graphicx,float}
\usepackage[T2A,T1]{fontenc}
\usepackage[utf8]{inputenc}
\usepackage[russian,english]{babel}

\usepackage{tipa}

\newcommand{\scell}[2][c]{%
  \begin{tabular}[#1]{@{}c@{}}#2\end{tabular}}
\aclfinalcopy % Uncomment this line for the final submission
\def\aclpaperid{304} %  Enter the acl Paper ID here

\title{A Crowdsourced Open-Source Kazakh Speech Corpus\\and Initial Speech Recognition Baseline}

\author{Yerbolat Khassanov, Saida Mussakhojayeva, Almas Mirzakhmetov,\\ \textbf{Alen Adiyev, Mukhamet Nurpeiissov and Huseyin Atakan Varol} \\
  Institute of Smart Systems and Artificial Intelligence (ISSAI),\\ Nazarbayev University, Nur-Sultan, Kazakhstan \\
  \texttt{\{yerbolat.khassanov, saida.mussakhojayeva, almas.mirzakhmetov,}\\
  \texttt{alen.adiyev, mukhamet.nurpeiissov, ahvarol\}@nu.edu.kz}\\}

\date{}

\begin{document}
\maketitle
\begin{abstract}
%Abstract should be no longer than 200 words.

We present an open-source speech corpus for the Kazakh language.
The Kazakh speech corpus (KSC) contains around 332 hours of transcribed audio comprising over 153,000 utterances spoken by participants from different regions and age groups, as well as both genders.
It was carefully inspected by native Kazakh speakers to ensure high quality.
The KSC is the largest publicly available database developed to advance various Kazakh speech and language processing applications.
In this paper, we first describe the data collection and preprocessing procedures followed by a description of the database specifications.
We also share our experience and challenges faced during the database construction, which might benefit other researchers planning to build a speech corpus for a low-resource language.
%The described methodologies will benefit other researchers planning to build speech corpus for low-resource language and are applicable during the the lockdown period.
To demonstrate the reliability of the database, we performed preliminary speech recognition experiments.
The experimental results imply that the quality of audio and transcripts is promising (2.8\% character error rate and 8.7\% word error rate on the test set).
To enable experiment reproducibility and ease the corpus usage, we also released an ESPnet recipe for our speech recognition models.

\end{abstract}

%##########################################################################################################################################
\section{Introduction}
%What is the aim of this paper?
We present an open-source Kazakh speech corpus (KSC) constructed to advance the development of speech and language processing applications for the Kazakh language.
Kazakh is an agglutinative language with vowel harmony and belongs to the family of Turkic languages.
During the Soviet period, the Kazakh language was overwhelmed by the Russian language, which caused a decline in Kazakh language usage~\citep{dave2007kazakhstan}.
In the 1990s, it was declared an official language of Kazakhstan, and many initiatives were launched to increase the number of Kazakh speakers.
Today, it is spoken by over 10 million people in Kazakhstan and by over 3 million people in other countries\footnote{\url{https://www.ethnologue.com/language/kaz}}.
By introducing the KSC, we aim to accelerate the penetration of the Kazakh language into the Internet of things (IoT) technologies and to promote research in Kazakh speech processing applications.

%What have been previously done to advance research for Kazakh language?
Although several Kazakh speech corpora have been presented~\citep{DBLP:conf/emnlp/MakhambetovMYMSS13,DBLP:conf/apsipa/ShiHTWZ17,DBLP:conf/aciids/MamyrbayevTMAKT19}, there is no generally accepted common corpus.
Most of them are either publicly unavailable or contain an insufficient amount of data to train reliable models.
Especially, these databases are too small for building recent end-to-end models, which are extremely data hungry~\citep{Hannun2014DeepSS}.
Consequently, different research groups usually conduct their experiments on internally collected data, which prevents the reproducibility and comparison of different approaches.

%What makes your work different from others?
To address the aforementioned limitations, we created the KSC, containing around 332 hours of transcribed audio.
It was crowdsourced through the Internet, where volunteers were asked to read sentences presented through a web browser.
In total, we accepted over 153,000 utterances submitted from over 1,600 unique device IDs. %identified by the web cookies stored on the users' devices.
%addresses\footnote{Due to the dynamic nature of some IP addresses, it is difficult to identify the number of unique speakers.}.
The recordings were first checked manually, and when a sufficient amount of data was collected, were partially checked automatically using a speech recognition system.
To the best of our knowledge, the KSC is the largest open-source speech corpus in Kazakh and is available for academic and commercial use upon request at this link\footnote{\url{https://issai.nu.edu.kz/kz-speech-corpus/?version=1.1}} under the Creative Commons Attribution 4.0 International License\footnote{\url{https://creativecommons.org/licenses/by/4.0/}}. %\footnote{\url{https://issai.nu.edu.kz/kz-speech-corpus/}}

We expect that this database will be a valuable resource for research communities in both academia and industry.
The primary application domains of the corpus are speech recognition, speech synthesis, and speaker recognition.
To demonstrate the reliability of the database, we performed preliminary automatic speech recognition (ASR) experiments, where promising and sufficient for practical usage results were achieved.
We also provide a practical guide on the development of ASR systems for the Kazakh language by sharing the reproducible recipe and pretrained models\footnote{\url{https://github.com/IS2AI/ISSAI_SAIDA_Kazakh_ASR}}. %\footnote{\url{https://github.com/IS2AI/ISSAI_SAIDA_Kazakh_ASR}}
The utilization of the database for speech synthesis and speaker recognition tasks is left for future work.

%How paper is organized?
The rest of the paper is organized as follows.
Section~\ref{sec:related_works} provides a review of related works.
Section~\ref{sec:ksc_data} presents the KSC database and explains the database construction procedures in detail.
Section~\ref{sec:experiment} offer a presentation of the speech recognition experiment setup and obtained results.
In Section~\ref{sec:discussion}, the obtained results and challenges are discussed.
Lastly, Section~\ref{sec:conclusion} concludes this paper and mentions potential future work.

%##########################################################################################################################################
\section{Related Works}
\label{sec:related_works}
%The ASR has intrigued researchers for centuries.
In the past few years, the interest in ASR has surged by its new applications in smart devices, such as voice command, voice search, message dictation, and virtual assistants~\citep{yu2016automatic}.
In response to this technological shift, many speech corpora have been introduced for various languages.
For example,~\citet{DBLP:journals/corr/abs-1808-10583} released 1,000 hours of open-source Mandarin read speech data to bridge the gap between academia and industry.
The utterances were recorded using iOS-based mobile phones and cover the following domains: voice command, smart home, autonomous driving, and so on. 
Similarly,~\citet{DBLP:conf/interspeech/KohMKAANT19} developed a 2,000-hour read speech corpus to help speech technology developers and researchers build speech-related applications in Singapore.

Several works have presented speech databases for the Kazakh language.
For example, \citet{DBLP:conf/emnlp/MakhambetovMYMSS13} developed a Kazakh language corpus containing around 40 hours of transcribed read speech data recorded in a sound-proof studio.
Similarly, \citet{DBLP:conf/aciids/MamyrbayevTMAKT19} collected 76 hours of data using a professional recording booth which were further extended to 123 hours in~\citet{DBLP:conf/aciids/MamyrbayevAZTG20}.
\citet{DBLP:conf/specom/KhomitsevichMTR15} utilized 147 hours of bilingual Kazakh-Russian speech data to build code-switching ASR systems.
\citet{DBLP:conf/apsipa/ShiHTWZ17} released 78 hours of transcribed Kazakh speech data recorded by 96 students from China.
The IARPA Babel project has released a Kazakh language pack\footnote{\url{https://catalog.ldc.upenn.edu/LDC2018S13}} consisting of around 50 hours of conversational and 14 hours of scripted telephone speech.
Unfortunately, the aforementioned databases are either publicly unavailable or of an insufficient size to build robust Kazakh ASR systems.
Additionally, some of them are nonrepresentative--that is, they cover speakers from a narrow set of categories, such as the same region or age group.   
Furthermore, since most of these databases have been collected in optimal lab settings, they might be ineffective for real-world applications.
%Furthermore, the research findings made on smaller databases can't reliably generalize to large data settings used in real scenarios~\cite{DBLP:journals/corr/abs-1808-10583}. 

The emergence of crowdsourcing platforms and the growth in Internet connectivity has motivated researchers to employ crowdsourcing for annotated corpora construction.
Different from the expert-based approaches, crowdsourcing tends to be cheaper and faster, though, additional measures should be taken to ensure the data quality~\cite{DBLP:conf/emnlp/SnowOJN08,DBLP:conf/naacl/NovotneyC10,eskenazi2013crowdsourcing}.
Furthermore, crowdsourcing allows to gather a variety of dialects and accents from remote geographical locations and enables the participation of people with disabilities and of an advanced age, which otherwise would be impossible or too costly~\citep{DBLP:conf/lrec/TakamichiS18}. 
Inspired by this, we followed the best crowdsourcing practices to construct a large-scale and open-source speech corpus for the Kazakh language, as described in the following sections.

%##########################################################################################################################################
\section{The KSC Construction}
\label{sec:ksc_data}
The KSC project was conducted with the approval of the Institutional Research Ethics Committee of Nazarbayev University.
Each reader participated voluntarily and was informed of the data collection and use protocols through an online consent form.

\subsection{Text Collection and Cleaning}
We first extracted Kazakh textual data from various sources such as electronic books, laws, and websites, including Wikipedia, news portals, and blogs.
For each website, we designed a specialized web crawler to improve the quality of the extracted text.
The extracted texts were manually filtered to eliminate inappropriate content involving sensitive political issues, user privacy, violence, and so on.
Additionally, we filtered out texts entirely consisting of Russian words.
Texts consisting of mixed Kazakh-Russian utterances were kept, because there are many borrowed Russian words in Kazakh, and it is common practice among Kazakh speakers to code-switch between Kazakh and Russian~\citep{DBLP:conf/specom/KhomitsevichMTR15}.
Next, we split the texts into sentences and removed sentences consisting of more than 25 words.
Lastly, duplicate sentences were removed.
The total number of extracted sentences was around 2.3 million.

\subsection{Text Narration and Checking}

\begin{figure}[b]
    \centering
    \frame{\includegraphics[width=0.87\linewidth,trim={6cm 0cm 6.65cm 0cm},clip=true]{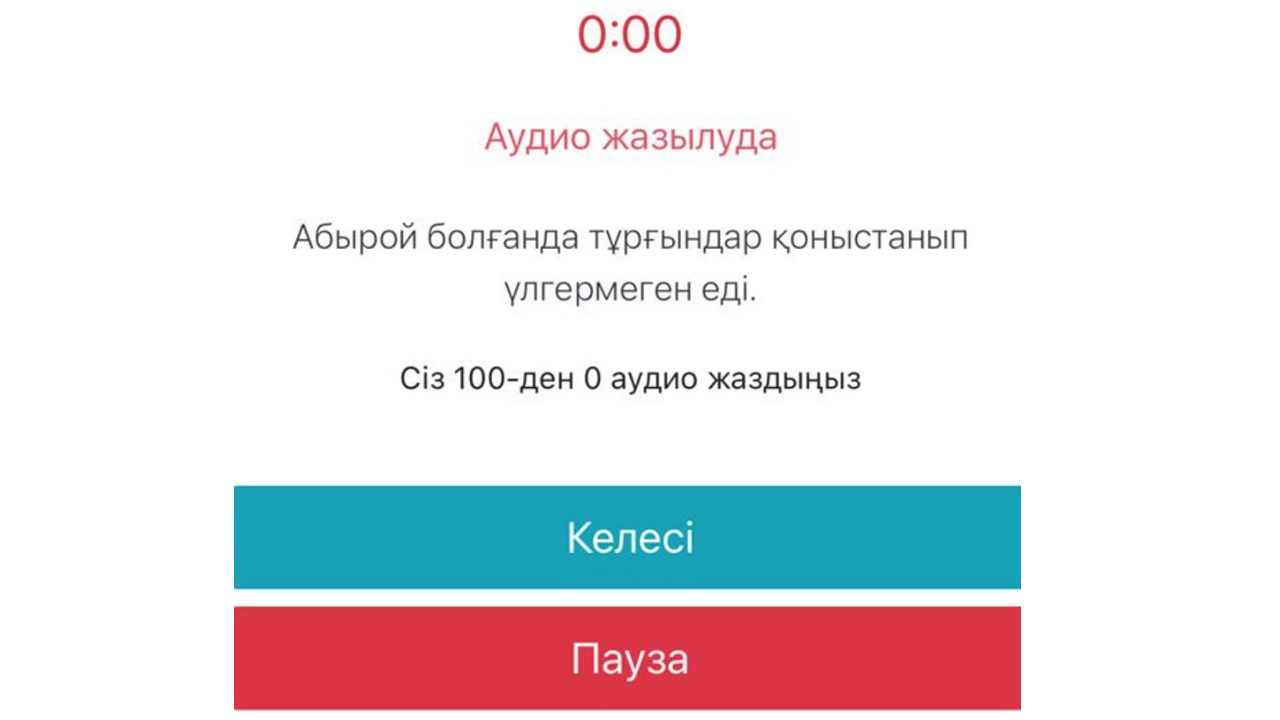}}
    \caption{\label{fig:record_page} The speech recording web interface.}
\end{figure}

To narrate the extracted sentences, we developed a web-based speech recording platform capable of running on personal computers and smartphones.
The platform randomly samples a sentence from the pool of extracted texts, and presents it to a reader (see Figure~\ref{fig:record_page}).
It also displays the recording status and statistics, such as elapsed time and the total number of read sentences.
Additionally, it has the ``pause'' and ``next'' buttons to control the recording process.
The readers were allowed to quit at any time.
We recruited readers by advertising the project in social media, news, and open messaging communities on WhatsApp and Telegram.
Readers who were at least 18 years old were included so that they could legally agree to participate in data collection.
The audios were recorded in 48 kHz and 16 bits, but downsampled to 16 kHz and 16 bits for online publication. 
%All recordings were saved as WAV files.
Following our experimental protocol, we did not store readers’ personal information except for the geolocation coordinates, IP address, and device type.
%Instead, each speaker was assigned a unique speaker ID which is released with the database.

Several native Kazakh transcribers were hired to check the quality of recordings.
The transcribers logged on to a special transcription checking platform and were provided with an audio segment and the corresponding sentence that a reader had read.
The task was to check if the reader had read the sentence according to the prompt, and to transcribe any deviations or other acoustic events based on a set of transcription instructions.
As an additional quality measure, we hired a linguist who was assigned to supervise the transcribers and to randomly check tasks they completed.
To harmonize the transcriptions, the linguist also held the ``go through errors'' sessions with the transcribers.

The transcribers were instructed to reject utterances containing obvious mispronunciations or severe noises, to convert numbers into words\footnote{Note that after converting numbers into words some sentence lengths exceeded 25 words.}, and to trim long silences at the beginning and end of the audio segments.
Additionally, they were instructed to enclose partial repetitions and hesitations in parentheses, for example, `(he) hello' and `(ah)', and to indicate other non-verbal sounds produced by readers, such as sneezing and coughing, using a special `[noise]' token. 
Background noises were not labeled.

When the size of accepted utterances reached 100 hours, we built an ASR system to automatically check the recordings.
The system accepted only recordings perfectly matching corresponding text prompts--that is, 0\% character error rate (CER), whereas the remaining utterances were left to human transcribers.

\subsection{Database Specifications}

\begin{table}[t]
    \small
    \renewcommand\arraystretch{1.1}
    \setlength{\tabcolsep}{1.5mm}
    \centering
    \begin{tabular}{l|c|c|c|c}
        \toprule
        Category                & Train         & Valid         & Test      & Total \\
        \midrule
        Duration (hours)        & 318.4         & 7.1           & 7.1       & 332.6 \\
        \# Utterances           & 147,236       & 3,283         & 3,334     & 153,853 \\
        %\# Words                &  1,614,620             &   36,095            &   36,431        &   \\
        \# Words                & 1.61M         & 35.2k         & 35.8k     & 1.68M \\
        \# Unique Words         & 157,191       & 13,525        & 13,959    & 160,041 \\
        \# Device IDs           & 1,554         & 29            & 29        & 1,612 \\
        \# Speakers             & -             & 29            & 29        & -  \\
        \bottomrule
    \end{tabular}
    \caption{\label{tab:stats}The KSC database specifications.}
\end{table}

The KSC database specifications are provided in Table~\ref{tab:stats}.
We split the data into three sets of non-overlapping speakers: training, validation, and test.
While the training set recordings were collected from anonymous speakers, the validation and test sets were collected from identifiable speakers to ensure that they did not overlap with the training set, represented different age groups and regions, and were gender balanced (see Table~\ref{tab:stats_dev_test}).
In total, around 153,000 utterances were accepted, yielding 332 hours of transcribed speech data.
%The distribution of utterances by regions is shown in Figure \textbf{X} and the distribution of utterances recorded by different speakers is shown in Figure \textbf{X}.
%The distribution of utterances recorded by different devices is shown in Figure~\ref{fig:utt_device}.
Note that the device IDs could not be used to represent the number of speakers in the training set as several speakers might have used the same device or the same speaker might have used different devices.
Therefore, the total number of speakers in the training set is not shown in Table~\ref{tab:stats}.
The whole database creation process took around four months, and the database size is around 38GB.

\begin{table}[t]
    \small
    \renewcommand\arraystretch{1.08}
    \setlength{\tabcolsep}{3mm}
    \centering
    \begin{tabular}{l|l|c|c}
        \toprule
        \multicolumn{2}{l|}{Category}                   & Valid         & Test      \\
        \midrule
        \multirow{2}{*}{Gender (\%)}    & Female        & 51.7          & 51.7      \\
                                        & Male          & 48.3          & 48.3      \\\hline
        \multirow{4}{*}{Age (\%)}       & 18-27         & 37.9          & 34.5      \\
                                        & 28-37         & 34.5          & 31.0      \\
                                        & 38-47         & 10.4          & 13.8      \\
                                        & 48 and above  & 17.2          & 20.7      \\\hline
        \multirow{5}{*}{Region (\%)}    & East          & 13.8          & 13.8      \\
                                        & West          & 20.7          & 17.2      \\
                                        & North         & 13.8          & 20.7      \\
                                        & South         & 37.9          & 41.4      \\
                                        & Center        & 13.8          & 6.9       \\\hline
        \multirow{2}{*}{Device (\%)}    & Phone         & 62.1          & 79.3      \\
                                        & Computer      & 37.9          & 20.7      \\\hline
        \multirow{2}{*}{Headphone (\%)} & Yes           & 20.7          & 17.2      \\
                                        & No            & 79.3          & 82.8      \\
        \bottomrule
    \end{tabular}
    \caption{\label{tab:stats_dev_test}The validation and test sets speaker details.}
\end{table}

\iffalse
\begin{figure}[b]
    \centering
    %\includegraphics[width=1\linewidth,height=9.5cm,trim={5.5cm 0cm 8.3cm 0.5cm},clip=true]{images/uttd_device_dist.png}
    \includegraphics[width=1\linewidth,trim={0cm 0cm 0cm 0cm},clip=true]{images/uttd_device_dist.png}
    \caption{\label{fig:utt_device} The distribution of utterances by devices in the training set of KSC.}
\end{figure}
\fi

\begin{figure}[t]
    \centering
    \includegraphics[width=0.9\linewidth,height=9.0cm,trim={5.65cm 0cm 8.4cm 0.5cm},clip=true]{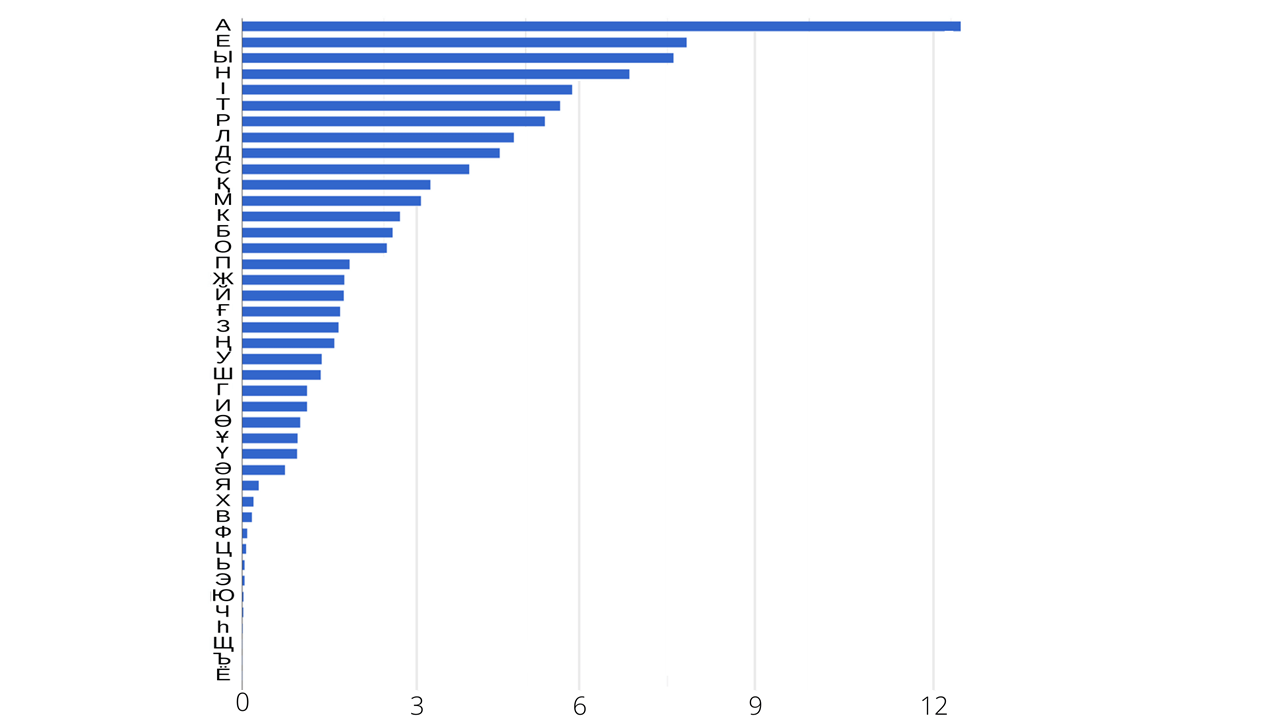}
    \caption{\label{fig:train_traj} The distribution of letters in the KSC (\%).}
\end{figure}

The Kazakh writing system differs depending on the regions where the language is spoken.
For example, the Cyrillic alphabet is used in Kazakhstan and Mongolia, while an Arabic-derived alphabet is used in China.
In the KSC, we presented all texts using the Cyrillic alphabet consisting of 42 letters.
The distribution of these letters in the KSC is given in Figure~\ref{fig:train_traj}.

One of the important features of the proposed KSC database is that it was collected in various environment conditions (e.g. home, office, café, transport, and street), with diverse background noises through mobile devices (e.g. phones and tablets) and personal computers, with and without headphone sets, which is similar to realistic use-case scenarios.
Consequently, our database enables the development and evaluation of ASR systems designed to operate in real-world voice-enabled applications, such as voice commands, voice search, message dictation, and so on.

%The KSC database structure is presented in Appendix \textbf{X} and it can be downloaded as a single file here\footnote{link}.
The KSC database consists of audio recordings, transcripts, and metadata stored in separate folders.
The audio and corresponding transcription filenames are the same, except that the audio recordings are stored as WAV files, whereas the transcriptions are stored as TXT files using the UTF-8 encoding.
The metadata contain the data splitting information (training, validation, and test) and the speaker's details (gender, age, region, device, and headphones) of the validation and test sets.

%##########################################################################################################################################
\section{Speech Recognition Experiment}
\label{sec:experiment}
To demonstrate the utility and reliability of the KSC database, we conducted speech recognition experiments using both the traditional deep neural network-hidden Markov model (DNN-HMM) and recently proposed end-to-end (E2E) architectures.
We did not compare or perform thorough architecture searches for either DNN-HMM and E2E, since this falls outside the scope of this paper.

\subsection{Experimental Setup}
All ASR models were trained with a single V100 GPU running on the NVIDIA DGX-2 server using the training set.
All hyper-parameters were tuned using the validation set. The best-performing model was evaluated using the test set.
All results are reported without lattice or n-best hypotheses rescoring, and no external data have been used.

\iffalse
%paper which studies the removal of disfluencies for spontaneous speech recognition
Reconstructing false start errors in spontaneous speech text
\fi
\subsubsection{The DNN-HMM ASR}
The DNN-HMM ASR system was built using the Kaldi framework~\citep{povey2011kaldi}.
We followed the Wall Street Journal (WSJ) recipe with the ``nnet3+chain'' setup and other latest Kaldi developments.
%The acoustic model was constructed by stacking 13 time-delay neural network (TDNN)~\citep{DBLP:conf/interspeech/PoveyCWLXYK18} layers with the dimension of 1,024 trained with the lattice-free maximum mutual information (LF-MMI) training criterion~\citep{DBLP:conf/interspeech/PoveyPGGMNWK16}.
The acoustic model was constructed using the factorized time-delay neural networks (TDNN-F)~\citep{DBLP:conf/interspeech/PoveyCWLXYK18} trained with the lattice-free maximum mutual information (LF-MMI)~\citep{DBLP:conf/interspeech/PoveyPGGMNWK16} training criterion.
The inputs were Mel-frequency cepstral coefficients (MFCC) features with cepstral mean and variance normalization extracted every 10 ms over a 25 ms window.
In addition, we applied data augmentation using the speed perturbation (SpeedPerturb)~\citep{DBLP:conf/interspeech/KoPPK15} at rates of 0.9, 1.0, and 1.1 and the spectral augmentation (SpecAugment)~\citep{DBLP:conf/interspeech/ParkCZCZCL19} techniques.

We employed a graphemic lexicon because of the strong correspondence between word spelling and pronunciation in Kazakh (e.g. ``$hello \rightarrow h\ e\ l\ l\ o$'').
The graphemic lexicon was constructed by extracting all words in the training set, which resulted in 157,191 unique words.
During the decoding stage, we employed a 3-gram language model\footnote{We tested different N-gram orders and found the 3-gram to perform best on our setup.} (LM) with the Kneser-Ney smoothing built using the SRILM toolkit~\citep{DBLP:conf/interspeech/Stolcke02}.
The 3-gram LM was trained using the transcripts of the training set and the vocabulary covering all the words in the graphemic lexicon.
%We didn't apply lattice or n-best list rescoring.

\subsubsection{The E2E ASR}
The E2E ASR systems were built using the \mbox{ESPnet} framework~\citep{watanabe2018espnet}.
We followed the WSJ recipe to train two different encoder-decoder architectures based on the recurrent neural networks (RNN)~\citep{DBLP:journals/corr/BahdanauCB14} and the Transformer~\citep{DBLP:conf/nips/VaswaniSPUJGKP17}.
Both architectures were jointly trained with the connectionist temporal classification (CTC)~\citep{DBLP:conf/icml/GravesFGS06} objective function under the multi-task learning framework~\citep{DBLP:conf/icassp/KimHW17}.
The input speech was represented as an 80-dimensional filterbank features with pitch computed every 10 ms over a 25 ms window.
For both E2E architectures, the acoustic features were first processed by few initial blocks of VGG network~\citep{DBLP:journals/corr/SimonyanZ14a}.
Since Kazakh is a morphologically rich language, it is susceptible to severe data sparseness.
To overcome this issue, we employed character-level output units in both architectures.
In total, we used 45 distinct output units consisting of 42 letters from the Kazakh alphabet and 3 special tokens--that is, $<$\textit{unk}$>$, $<$\textit{space}$>$, and $<$\textit{blank}$>$ used in CTC.

The E2E ASR systems do not require a lexicon when modeling with grapheme-based output units~\citep{DBLP:conf/icassp/SainathPKLKRSNL18}.
The character-level LM was built using the transcripts of the training set as a two-layer RNN with 650 long short-term memory (LSTM)~\citep{DBLP:journals/neco/HochreiterS97} units each.
We utilized the LSTM LM during the decoding stage using the shallow fusion~\citep{DBLP:journals/corr/GulcehreFXCBLBS15} for both E2E architectures.
%Additionally, the word-level RNN LM also trained using the transcripts of the train set was employed to rescore 50-best output hypotheses.
Besides, we augment the training data using the SpeedPerturb and the SpecAugment techniques.
For decoding, we set the beam size to 30 and the LSTM LM interpolation weight to 1.

\textbf{E2E-RNN.}
The encoder module of the RNN-based E2E ASR system consists of three bi-directional LSTM layers with 1,024 units per direction per layer.
The decoder module is a single uni-directional LSTM with 1,024 units.
We train the model for 20 epochs using the Adadelta optimizer with the initial learning rate set to 1 and the batch size set to 30.
The interpolation weight for the CTC objective was set to 0.5.

\textbf{E2E-Transformer.}
The Transformer-based E2E ASR system consists of 12 encoder and 6 decoder blocks.
We set the number of heads in the self-attention layer to 4 with 256-dimension hidden states and the feed-forward network dimensions to 2,048.
%We used the dropout rate of 0.1 and the label smoothing of 0.1.
We set the dropout rate and label smoothing to 0.1.
The model was trained for 160 epochs using the Noam optimizer~\cite{DBLP:conf/nips/VaswaniSPUJGKP17} with the initial learning rate of 10 and the warmup-steps of 25,000.
The batch size was set to 96.
We report results on an average model constructed using the last 10 checkpoints.
The interpolation weight for the CTC objective was set to 0.3.
%To the best of our knowledge, it is the first time the Transformer based ASR systems are build for the Kazakh language. 
%The ``drop-in and run'' ESPnet recipe is provided\footnote{\url{link_to_recipe}}.

\begin{table*}[h]
    %\small
    \renewcommand\arraystretch{1.1}
    \setlength{\tabcolsep}{2mm}
    \centering
    \begin{tabular}{c|l|c|c|c|c|c|c|c}
        \toprule
        \multirow{2}{*}{ID} & \multirow{2}{*}{Model}            & \multirow{2}{*}{LM}   & \multirow{2}{*}{SpeedPerturb} & \multirow{2}{*}{SpecAugment}  & \multicolumn{2}{|c}{Valid}& \multicolumn{2}{|c}{Test} \\\cline{6-9}
                            &                                   &                       &                               &                               & CER               & WER           & CER           & WER   \\
        \midrule
        1                   & \multirow{2}{*}{DNN-HMM}          & Yes                   & Yes                           & No                            & 5.2               & 14.2          & 4.6           & 13.7  \\
        2                   &                                   & Yes                   & Yes                           & Yes                           & 5.3               & 14.9          & 4.7           & 13.8  \\\hline
        3                   & \multirow{4}{*}{E2E-LSTM}         & No                    & No                            & No                            & 9.9               & 32.0          & 8.7           & 28.8    \\
        4                   &                                   & Yes                   & No                            & No                            & 7.9               & 20.1          & 7.2           & 18.5  \\
        5                   &                                   & Yes                   & Yes                           & No                            & 5.7               & 15.9          & 5.0           & 14.4  \\
        6                   &                                   & Yes                   & Yes                           & Yes                           & 4.6               & 13.1          & 4.0           & 11.7  \\\hline
        7                   & \multirow{4}{*}{E2E-Transformer}  & No                    & No                            & No                            & 6.1               & 22.2          & 4.9           & 18.8  \\
        8                   &                                   & Yes                   & No                            & No                            & 4.5               & 13.9          & 3.7           & 11.9  \\
        9                   &                                   & Yes                   & Yes                           & No                            & 3.9               & 12.3          & 3.2           & 10.5  \\
        10                  &                                   & Yes                   & Yes                           & Yes                           & \textbf{3.2}      & \textbf{10.0} & \textbf{2.8}  & \textbf{8.7}  \\
        \bottomrule
    \end{tabular}
    \caption{\label{tab:results}The CER (\%) and WER (\%) performances of different ASR models built using KSC.}
\end{table*}

\subsection{Experimental Results}
The experimental results are presented in Table~\ref{tab:results} in terms of both the character error rate (CER) and word error rate (WER).
All the ASR models achieved competitive results on both the validation and test sets.
We found that the validation set is more challenging than the test set.
%We also observed that the SpecAugment is ineffective for the DNN-HMM model despite trying different hyper-parameters.
When compared without SpecAugment, the performance of the DNN-HMM model (ID 1) is slightly better than the E2E-RNN (ID 5), but inferior to the E2E-Transformer (ID 9).
We could not achieve any improvements on the DNN-HMM model (ID 2) using the SpecAugment despite trying different hyper-parameter tuning recommendations~\cite{DBLP:conf/icassp/ZhouMIKSN20}. 
Overall, the best CER and WER results are achieved by the E2E-Transformer (ID 10) followed by the E2E-RNN (ID 6) and then the DNN-HMM (ID 1).

We observed that the LM fusion significantly improves the performances of both E2E models.
For example, 35\% and 36\% relative WER improvements are achieved on the test set for the RNN (from ID 3 to ID 4) and Transformer (from ID 7 to ID 8) models, respectively.
Furthermore, both data augmentation techniques based on SpeedPerturb and SpecAugment are highly effective for the Kazakh E2E ASR where additional improvements are achieved. 
For example, when models without and with data augmentations are compared, 36\% and 26\% relative WER improvements are achieved on the test set for the RNN (from ID 4 to ID 6) and Transformer (from ID 8 to ID 10) models, respectively.

These experimental results successfully demonstrate the utility of the KSC database for the speech recognition task.
We leave the exploration of the optimal hyper-parameter settings and detailed comparison of different ASR architectures for future work.

%##########################################################################################################################################

\section{Discussion}
\label{sec:discussion}

\begin{figure}[hb]
    \centering
    \includegraphics[width=1.0\linewidth,trim={5.75cm 1.75cm 6.25cm 1.5cm},clip=true]{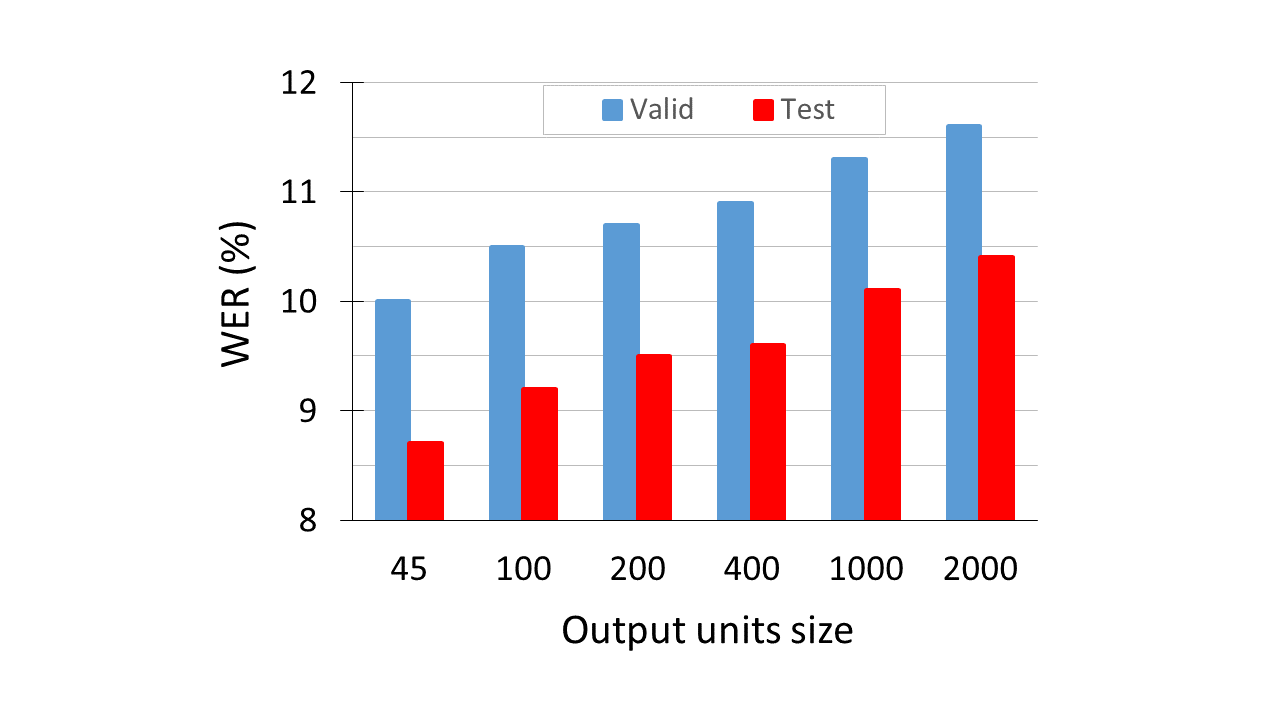}
    \caption{\label{fig:output_units} The impact of different output unit sizes on the E2E-Transformer ASR performance.}
\end{figure}

\textit{1) Data sparsity.}
Kazakh language speech recognition is considered challenging due to the agglutinative nature of the language, where word structures are formed by adding derivational and inflectional affixes to stems in a specific order.
As a result, the vocabulary size might considerably grow resulting in a data sparsity problem, especially for models operating in a word level, such as our DNN-HMM architecture.
The potential solution is to break down words into finer-level linguistic units, such as characters or subword units~\citep{DBLP:conf/acl/SennrichHB16a}.
%This will result in a huge vocabulary size leading to the data sparsity issues, especially for the models that operate in a word-level.
We investigated the impact of other output unit sizes on the performance of the Kazakh E2E-Transformer ASR and did not observe any considerable improvements over the character-level outputs (see Figure~\ref{fig:output_units}).
The output units were generated using the byte pair encoding (BPE) algorithm implemented in the SentencePiece~\cite{DBLP:conf/emnlp/KudoR18} tokenizer.

\begin{table*}[t]
    %\small
    %\renewcommand\arraystretch{1.0}
    \setlength{\tabcolsep}{1.5mm}
    \centering
    \begin{tabular}{c|c|c|c|c}
        \toprule
        ID                  & Confusion pairs                                                                       & Insertion                                             
                            & Deletion                                                                              & Substitution \\
        \midrule
        \multirow{3}{*}{1}  & \foreignlanguage{russian}{жоғарғы} $\Rightarrow$ \foreignlanguage{russian}{жоғары}    & \foreignlanguage{russian}{мен}                                    
                            & \foreignlanguage{russian}{де}                                                         & \foreignlanguage{russian}{да} \\
                            & (jogargy $\Rightarrow$ jogary)                                                        & (men)                                                                 
                            & (de)                                                                                  & (da) \\
                            & \scell{“upper” or “higher” $\Rightarrow$\\“top” or “high”}                            & “I” or “with”                                                         
                            & \scell{“too” (after a\\thin-vowelled word)}                                           & \scell{“too” (after a\\thick-voweled word)} \\\hline
                            
        \multirow{3}{*}{2}  & \foreignlanguage{russian}{өзеннің} $\Rightarrow$ \foreignlanguage{russian}{өзенінің}  & \foreignlanguage{russian}{бұл}                                    
                            & \foreignlanguage{russian}{ал}                                                         & \foreignlanguage{russian}{де} \\
                            & (uzennin $\Rightarrow$ uzeninin)                                                      & (bul)                                                             
                            & (al)                                                                                  & (de) \\
                            & \scell{“of a river” $\Rightarrow$\\“of the river”}                                    & “this” or “these”                                                        
                            & \scell{an imperative of\\“take” or “whereas”}                                         & \scell{“too” (after a\\thin-voweled word)} \\\hline
                            
        \multirow{3}{*}{3}  & \foreignlanguage{russian}{ас} $\Rightarrow$ \foreignlanguage{russian}{ақ}             & \foreignlanguage{russian}{бір}                                    
                            & \foreignlanguage{russian}{да}                                                         & \foreignlanguage{russian}{мен} \\
                            & (as $\Rightarrow$ aq)                                                                 & (bir)                                                                 
                            & (da)                                                                                  & (men) \\
                            & “meal” $\Rightarrow$ “white”                                                          & “one"                                                                 
                            & \scell{“too” (after a\\thick-vowelled word)}                                           & “I” or “with” \\\hline
                            
        \multirow{3}{*}{4}  & \foreignlanguage{russian}{бағы} $\Rightarrow$ \foreignlanguage{russian}{баға}         & \foreignlanguage{russian}{да}                                     
                            & \foreignlanguage{russian}{әр}                                                         & \foreignlanguage{russian}{жылы} \\
                            & (bagy $\Rightarrow$ baga)                                                             & (da)                                                              
                            & (ar)                                                                                  & (jyly) \\
                            & \scell{“the garden” $\Rightarrow$ “price”}                                            & \scell{“too” (after a\\thick-vowelled word)}                           
                            & \scell{“every”}                                                                       & \scell{“the year” or "warm"} \\\hline
                            
        \multirow{3}{*}{5}  & \foreignlanguage{russian}{болу} $\Rightarrow$ \foreignlanguage{russian}{болуы}        & \foreignlanguage{russian}{ақ}                             
                            & \foreignlanguage{russian}{бір}                                                         & \foreignlanguage{russian}{бір} \\
                            & (bolu $\Rightarrow$ boluy)                                                            & (aq)                                                                  
                            & (bir)                                                                                 & (bir) \\
                            & \scell{“being” $\Rightarrow$ “the being”}                                             & \scell{“white”}                                                       
                            & \scell{“one"}                                                                         & \scell{“one”} \\

        \bottomrule
    \end{tabular}
    \caption{\label{tab:output}The top five confusion pairs (reference word $\Rightarrow$ recognized word) and insertion, deletion, and substitution errors in the recognized output of the E2E-Transformer ASR.}
\end{table*}

\textit{2) Code-switching.}
Another challenge is the Kazakh-Russian code-switching practice which is common in daily communication as the majority of Kazakhs are bilingual.
Mostly, inter-sentential and intra-sentential types of code-switching are practiced, however, intra-word code-switching is also possible.
For example, one can say ``\textit{Men magazinge bardym}'' (``\textit{I went to a store}''), where the Russian word ``\textit{magazin}'' is appended by the Kazakh inflection ``\textit{-ge}'' representing the preposition ``\textit{to}''.
Furthermore, while the spelling of Kazakh words closely matches their pronunciation, this is not the case for Russian words, for example, the letter ``\textit{o}'' is sometimes pronounced as \textipa{/a/}, which might confuse an ASR system. %Show the alphabet confusion matrix
We observed that our ASR system is ineffective in code-switched utterances. 
Therefore, future work should focus on alleviating these errors.
%We inspected the recognized output results and observed that the ASR consistently makes errors on the code-mixed utterances and the rare proper nouns.

\begin{figure}[b!]
    \centering
    \includegraphics[width=1.0\linewidth,trim={7cm 1.75cm 6.5cm 2.5cm},clip=true]{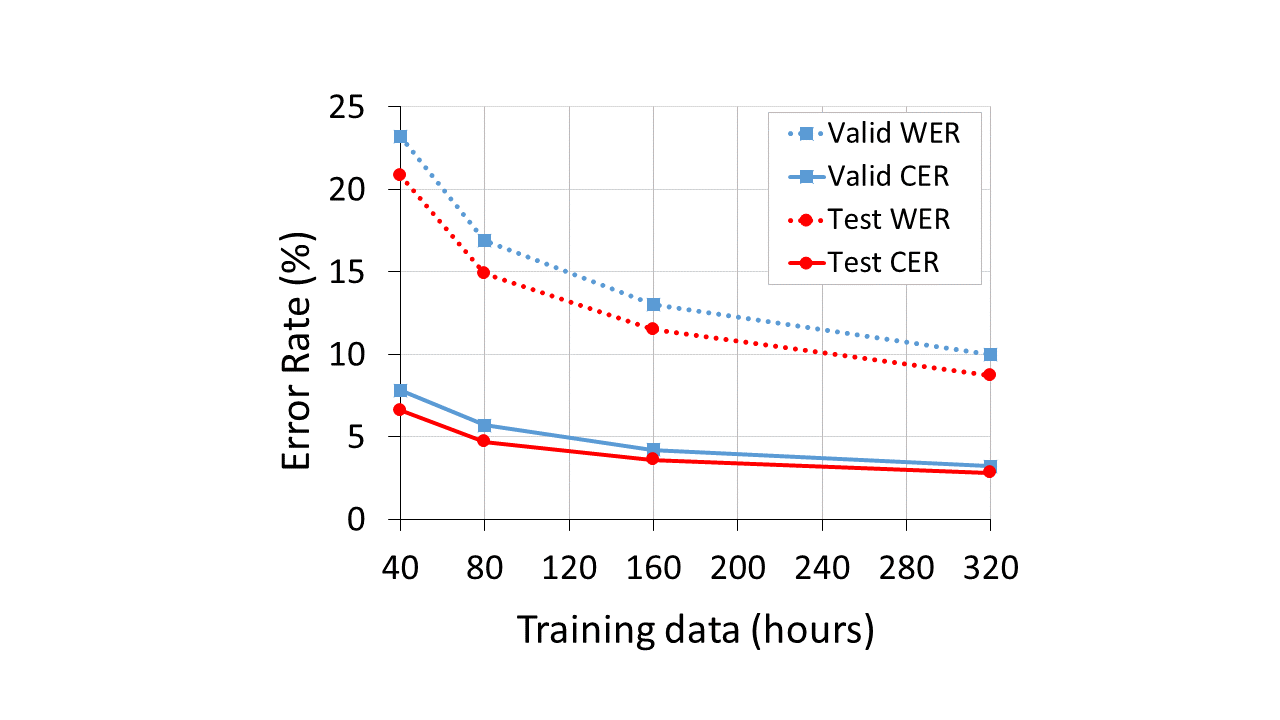}
    \caption{\label{fig:data_eff} The data efficiency experiment.}
\end{figure}

\textit{3) Data efficiency.}
To analyze the data efficiency--that is, increase in performance due to the addition of new training data--we trained our best E2E-Transformer ASR system using different amounts of data.
In particular, we first randomly sampled 40 hours of data and kept increasing their size until the entire training set was covered.
The experimental results indicate that the WER performance has not converged yet and further data collection might be effective (see Figure~\ref{fig:data_eff}).
Therefore, we plan to continue the data collection process.

\begin{figure*}[h]
    \centering
    \includegraphics[width=0.825\linewidth,trim={4.5cm 0cm 6.5cm 0cm},clip=true]{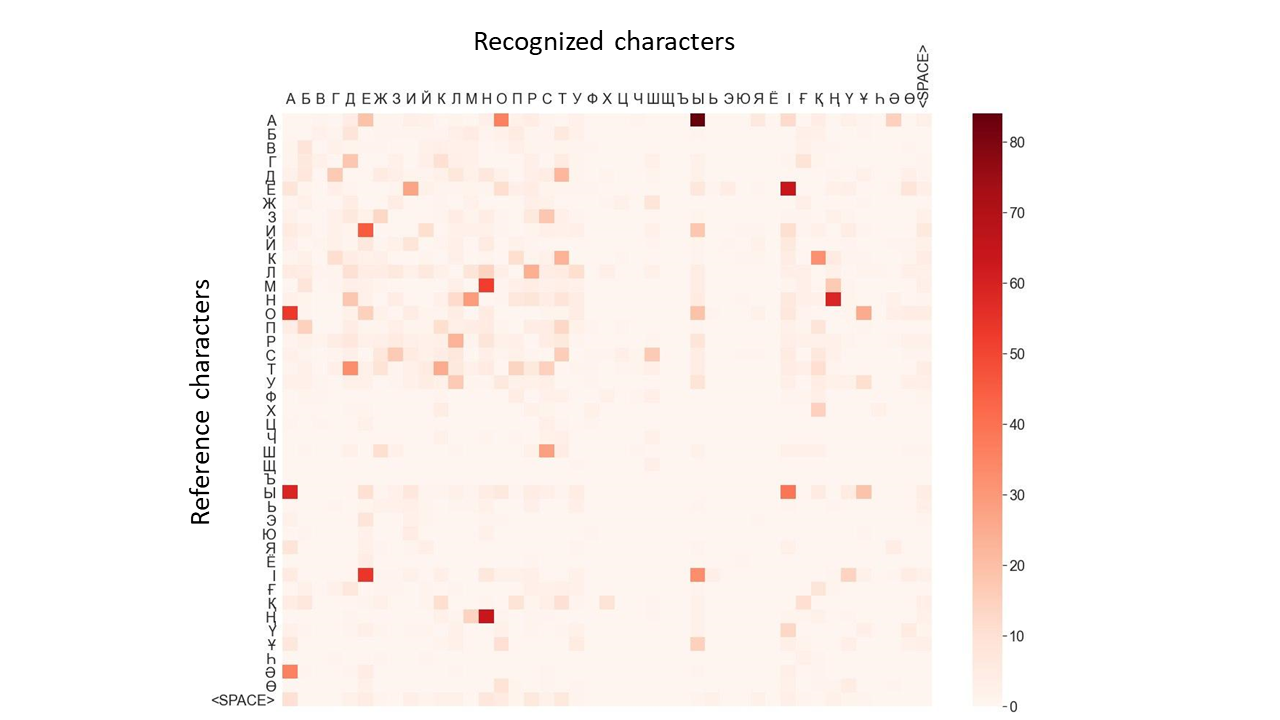}
    \caption{\label{fig:char_conf_mat} The character confusion matrix for the E2E-Transformer ASR.}
\end{figure*}

\textit{4) ASR output analysis.} 
We inspected the recognized outputs of the best E2E-Transformer model to identify the most challenging characters and words.
Table~\ref{tab:output} lists the top five confusion pairs and insertion, deletion, and substitution errors.
Most of them are commonly used words, such as conjunctions and numbers.
In addition, we also inspected the most confused character pairs (see Figure~\ref{fig:char_conf_mat}).
We observed that the Kazakh ASR system confuses characters with a similar pronunciation, such as ``\foreignlanguage{russian}{н}'' (\textipa{/n/}) and ``\foreignlanguage{russian}{ң}'' (\textipa{/\ng/}), ``\foreignlanguage{russian}{і}'' (\textipa{/\textsci/}) and ``\foreignlanguage{russian}{ы}'' (\textipa{/\textschwa/}), and so on. 

\textit{5) Performance comparison.}
%To alleviate the aforementioned challenges, a large amount of transcribed speech data on par with the effective speech recognition approaches are required. 
%In this regard, we developed the KSC database which will be a valuable tool for advancing Kazakh speech processing applications.
We cannot directly compare our results to previous works; however, our WER results are appealing.
For example,~\citet{DBLP:conf/aciids/MamyrbayevTMAKT19} used 76 hours of speech data to build an DNN-HMM ASR system which achieved 32.7\% WER on clean read speech.
Similarly, the DNN-HMM ASR system built using 78 hours of data in~\citep{DBLP:conf/apsipa/ShiHTWZ17} achieved 25.1\% WER on read speech.
On the other hand,~\citet{DBLP:conf/aciids/MamyrbayevAZTG20} achieved 17.8\% WER on clean read speech using the E2E ASR system trained on 126 hours of data. In comparison, our best model achieved 8.7\% WER on the test set.

\textit{6) Benefit for other Turkic Languages.}
We also envision that the KSC can be utilized in cross-lingual transfer learning techniques~\citep{DBLP:conf/interspeech/DasH15} to improve ASR systems for other Turkic languages, such as Kyrgyz and Tatar.

\iffalse
\begin{table}[h]
    \renewcommand\arraystretch{1.2}
    \setlength{\tabcolsep}{1.5mm}
    \centering
    \begin{tabular}{p{3.5cm}|p{3.5cm}}
    %\begin{tabular}{l|p{5.8cm}}
        \toprule
        Output              & Reference \\
        \midrule
        \textit{\small көрмедегі ең әдемі ең қымбат көліктер осы залда} &  көрмедегі ең әдемі ең қымбат көліктер осы залда  \\
        \bottomrule
    \end{tabular}
    \caption{\label{tab:examples}Output recognition examples of Transformer-based ASR model (ID 9).}
\end{table}
\fi

%##########################################################################################################################################
\section{Conclusion}
\label{sec:conclusion}
In this work, we presented the KSC database containing around 332 hours of transcribed speech data.
It was developed to advance Kazakh speech processing applications, such as speech recognition, speech synthesis, and speaker recognition.
We described the database construction procedures and discussed challenges that should be addressed in future work.
The described methodologies might benefit other researchers planning to build a speech corpus for a low-resource language.
The database is freely available for any purposes including research and commercial use.
We also conducted preliminary speech recognition experiments using both traditional hybrid DNN-HMM and recently proposed E2E ASR architectures.
To ease the database usage and ensure the reproducibility of experiments, we split it into three non-overlapping sets (training, validation, and test) and released our ESPnet recipe.
The detailed exploration of better ASR settings, as well as the adaptation of the database to other applications, is left for future work.

%##########################################################################################################################################
%\iffalse 
%Acknowledgement should be hidden during the reviewing process
\section*{Acknowledgments}
We would like to thank Aigerim Boranbayeva for helping to train and monitor the transcribers.
We also thank Kuralay Baimenova, Yerbol Absalyamov, Gibrat Kurmanov, and Rustem Yeshpanov for helping to attract readers and to resolve other administrative tasks.
%\fi

%##########################################################################################################################################
\bibliography{anthology,eacl2021}

\begin{thebibliography}{33}
\expandafter\ifx\csname natexlab\endcsname\relax\def\natexlab#1{#1}\fi

\bibitem[{Bahdanau et~al.(2015)Bahdanau, Cho, and
  Bengio}]{DBLP:journals/corr/BahdanauCB14}
Dzmitry Bahdanau, Kyunghyun Cho, and Yoshua Bengio. 2015.
\newblock \href {http://arxiv.org/abs/1409.0473} {Neural machine translation by
  jointly learning to align and translate}.
\newblock In \emph{Proc. of the 3rd International Conference on Learning
  Representations (ICLR), San Diego, CA, USA, May 7-9, 2015}.

\bibitem[{Das and Hasegawa{-}Johnson(2015)}]{DBLP:conf/interspeech/DasH15}
Amit Das and Mark Hasegawa{-}Johnson. 2015.
\newblock \href
  {http://www.isca-speech.org/archive/interspeech\_2015/i15\_3531.html}
  {Cross-lingual transfer learning during supervised training in low resource
  scenarios}.
\newblock In \emph{Proc. of the 16th Annual Conference of the International
  Speech Communication Association (INTERSPEECH), Dresden, Germany, September
  6-10, 2015}, pages 3531--3535.

\bibitem[{Dave(2007)}]{dave2007kazakhstan}
Bhavna Dave. 2007.
\newblock \emph{Kazakhstan-ethnicity, language and power}.
\newblock Routledge.

\bibitem[{Du et~al.(2018)Du, Na, Liu, and
  Bu}]{DBLP:journals/corr/abs-1808-10583}
Jiayu Du, Xingyu Na, Xuechen Liu, and Hui Bu. 2018.
\newblock \href {http://arxiv.org/abs/1808.10583} {{AISHELL-2:} transforming
  mandarin {ASR} research into industrial scale}.
\newblock \emph{CoRR}, abs/1808.10583.

\bibitem[{Eskenazi et~al.(2013)Eskenazi, Levow, Meng, Parent, and
  Suendermann}]{eskenazi2013crowdsourcing}
Maxine Eskenazi, Gina-Anne Levow, Helen Meng, Gabriel Parent, and David
  Suendermann. 2013.
\newblock \emph{Crowdsourcing for speech processing: Applications to data
  collection, transcription and assessment}.
\newblock John Wiley \& Sons.

\bibitem[{Graves et~al.(2006)Graves, Fern{\'{a}}ndez, Gomez, and
  Schmidhuber}]{DBLP:conf/icml/GravesFGS06}
Alex Graves, Santiago Fern{\'{a}}ndez, Faustino~J. Gomez, and J{\"{u}}rgen
  Schmidhuber. 2006.
\newblock \href {https://doi.org/10.1145/1143844.1143891} {Connectionist
  temporal classification: labelling unsegmented sequence data with recurrent
  neural networks}.
\newblock In \emph{Proc. of the 23rd International Conference on Machine
  Learning, {(ICML}), Pittsburgh, Pennsylvania, USA, June 25-29, 2006}, pages
  369--376.

\bibitem[{G{\"{u}}l{\c{c}}ehre et~al.(2015)G{\"{u}}l{\c{c}}ehre, Firat, Xu,
  Cho, Barrault, Lin, Bougares, Schwenk, and
  Bengio}]{DBLP:journals/corr/GulcehreFXCBLBS15}
{\c{C}}aglar G{\"{u}}l{\c{c}}ehre, Orhan Firat, Kelvin Xu, Kyunghyun Cho,
  Lo{\"{\i}}c Barrault, Huei{-}Chi Lin, Fethi Bougares, Holger Schwenk, and
  Yoshua Bengio. 2015.
\newblock \href {http://arxiv.org/abs/1503.03535} {On using monolingual corpora
  in neural machine translation}.
\newblock \emph{CoRR}, abs/1503.03535.

\bibitem[{Hannun et~al.(2014)Hannun, Case, Casper, Catanzaro, Diamos, Elsen,
  Prenger, Satheesh, Sengupta, Coates, and Ng}]{Hannun2014DeepSS}
Awni~Y. Hannun, C.~Case, J.~Casper, Bryan Catanzaro, Greg Diamos, E.~Elsen,
  Ryan Prenger, S.~Satheesh, S.~Sengupta, A.~Coates, and A.~Ng. 2014.
\newblock Deep speech: Scaling up end-to-end speech recognition.
\newblock \emph{ArXiv}, abs/1412.5567.

\bibitem[{Hochreiter and Schmidhuber(1997)}]{DBLP:journals/neco/HochreiterS97}
Sepp Hochreiter and J{\"{u}}rgen Schmidhuber. 1997.
\newblock \href {https://doi.org/10.1162/neco.1997.9.8.1735} {Long short-term
  memory}.
\newblock \emph{Neural Comput.}, 9(8):1735--1780.

\bibitem[{Khomitsevich et~al.(2015)Khomitsevich, Mendelev, Tomashenko, Rybin,
  Medennikov, and Kudubayeva}]{DBLP:conf/specom/KhomitsevichMTR15}
Olga Khomitsevich, Valentin Mendelev, Natalia~A. Tomashenko, Sergey Rybin, Ivan
  Medennikov, and Saule Kudubayeva. 2015.
\newblock \href {https://doi.org/10.1007/978-3-319-23132-7\_3} {A bilingual
  kazakh-russian system for automatic speech recognition and synthesis}.
\newblock In \emph{Proc. of the 17th International Conference on Speech and
  Computer (SPECOM), Athens, Greece, September 20-24, 2015}, volume 9319, pages
  25--33.

\bibitem[{Kim et~al.(2017)Kim, Hori, and Watanabe}]{DBLP:conf/icassp/KimHW17}
Suyoun Kim, Takaaki Hori, and Shinji Watanabe. 2017.
\newblock \href {https://doi.org/10.1109/ICASSP.2017.7953075} {Joint
  ctc-attention based end-to-end speech recognition using multi-task learning}.
\newblock In \emph{Proc. of the {IEEE} International Conference on Acoustics,
  Speech and Signal Processing (ICASSP), New Orleans, LA, USA, March 5-9,
  2017}, pages 4835--4839.

\bibitem[{Ko et~al.(2015)Ko, Peddinti, Povey, and
  Khudanpur}]{DBLP:conf/interspeech/KoPPK15}
Tom Ko, Vijayaditya Peddinti, Daniel Povey, and Sanjeev Khudanpur. 2015.
\newblock \href
  {http://www.isca-speech.org/archive/interspeech\_2015/i15\_3586.html} {Audio
  augmentation for speech recognition}.
\newblock In \emph{Proc. of the 16th Annual Conference of the International
  Speech Communication Association ({INTERSPEECH}), Dresden, Germany, September
  6-10, 2015}, pages 3586--3589.

\bibitem[{Koh et~al.(2019)Koh, Mislan, Khoo, Ang, Ang, Ng, and
  Tan}]{DBLP:conf/interspeech/KohMKAANT19}
Jia~Xin Koh, Aqilah Mislan, Kevin Khoo, Brian Ang, Wilson Ang, Charmaine Ng,
  and Ying{-}Ying Tan. 2019.
\newblock \href {https://doi.org/10.21437/Interspeech.2019-1525} {Building the
  singapore english national speech corpus}.
\newblock In \emph{Proc. of the 20th Annual Conference of the International
  Speech Communication Association (INTERSPEECH), Graz, Austria, 15-19
  September 2019}, pages 321--325.

\bibitem[{Kudo and Richardson(2018)}]{DBLP:conf/emnlp/KudoR18}
Taku Kudo and John Richardson. 2018.
\newblock \href {https://doi.org/10.18653/v1/d18-2012} {Sentencepiece: {A}
  simple and language independent subword tokenizer and detokenizer for neural
  text processing}.
\newblock In \emph{Proceedings of the 2018 Conference on Empirical Methods in
  Natural Language Processing, {EMNLP} 2018: System Demonstrations, Brussels,
  Belgium, October 31 - November 4, 2018}, pages 66--71. Association for
  Computational Linguistics.

\bibitem[{Makhambetov et~al.(2013)Makhambetov, Makazhanov, Yessenbayev,
  Matkarimov, Sabyrgaliyev, and
  Sharafudinov}]{DBLP:conf/emnlp/MakhambetovMYMSS13}
Olzhas Makhambetov, Aibek Makazhanov, Zhandos Yessenbayev, Bakhyt Matkarimov,
  Islam Sabyrgaliyev, and Anuar Sharafudinov. 2013.
\newblock \href {https://www.aclweb.org/anthology/D13-1104/} {Assembling the
  kazakh language corpus}.
\newblock In \emph{Proc. of the Conference on Empirical Methods in Natural
  Language Processing (EMNLP), 18-21 October 2013, Seattle, Washington, USA},
  pages 1022--1031.

\bibitem[{Mamyrbayev et~al.(2020)Mamyrbayev, Alimhan, Zhumazhanov, Turdalykyzy,
  and Gusmanova}]{DBLP:conf/aciids/MamyrbayevAZTG20}
Orken Mamyrbayev, Keylan Alimhan, Bagashar Zhumazhanov, Tolganay Turdalykyzy,
  and Farida Gusmanova. 2020.
\newblock \href {https://doi.org/10.1007/978-3-030-42058-1\_33} {End-to-end
  speech recognition in agglutinative languages}.
\newblock In \emph{Proc. of the 12th Asian Conference on Intelligent
  Information and Database Systems (ACIIDS), Phuket, Thailand, March 23-26,
  2020}, volume 12034, pages 391--401.

\bibitem[{Mamyrbayev et~al.(2019)Mamyrbayev, Turdalyuly, Mekebayev, Alimhan,
  Kydyrbekova, and Turdalykyzy}]{DBLP:conf/aciids/MamyrbayevTMAKT19}
Orken~J. Mamyrbayev, Mussa Turdalyuly, Nurbapa Mekebayev, Keylan Alimhan, Aizat
  Kydyrbekova, and Tolganay Turdalykyzy. 2019.
\newblock \href {https://doi.org/10.1007/978-3-030-14802-7\_40} {Automatic
  recognition of kazakh speech using deep neural networks}.
\newblock In \emph{Proc. of the 11th Asian Conference on Intelligent
  Information and Database Systems (ACIIDS), Yogyakarta, Indonesia, April 8-11,
  2019}, pages 465--474.

\bibitem[{Novotney and Callison{-}Burch(2010)}]{DBLP:conf/naacl/NovotneyC10}
Scott Novotney and Chris Callison{-}Burch. 2010.
\newblock \href {https://www.aclweb.org/anthology/N10-1024/} {Cheap, fast and
  good enough: Automatic speech recognition with non-expert transcription}.
\newblock In \emph{Proc. of Human Language Technologies: Conference of the
  North American Chapter of the Association of Computational Linguistics, June
  2-4, 2010, Los Angeles, California, {USA}}, pages 207--215.

\bibitem[{Park et~al.(2019)Park, Chan, Zhang, Chiu, Zoph, Cubuk, and
  Le}]{DBLP:conf/interspeech/ParkCZCZCL19}
Daniel~S. Park, William Chan, Yu~Zhang, Chung{-}Cheng Chiu, Barret Zoph,
  Ekin~D. Cubuk, and Quoc~V. Le. 2019.
\newblock \href {https://doi.org/10.21437/Interspeech.2019-2680} {Specaugment:
  {A} simple data augmentation method for automatic speech recognition}.
\newblock In \emph{Proc. of the 20th Annual Conference of the International
  Speech Communication Association (INTERSPEECH), Graz, Austria, 15-19
  September 2019}, pages 2613--2617.

\bibitem[{Povey et~al.(2018)Povey, Cheng, Wang, Li, Xu, Yarmohammadi, and
  Khudanpur}]{DBLP:conf/interspeech/PoveyCWLXYK18}
Daniel Povey, Gaofeng Cheng, Yiming Wang, Ke~Li, Hainan Xu, Mahsa Yarmohammadi,
  and Sanjeev Khudanpur. 2018.
\newblock \href {https://doi.org/10.21437/Interspeech.2018-1417}
  {Semi-orthogonal low-rank matrix factorization for deep neural networks}.
\newblock In \emph{Proc. of the 19th Annual Conference of the International
  Speech Communication Association (INTERSPEECH), Hyderabad, India, 2-6
  September 2018}, pages 3743--3747.

\bibitem[{Povey et~al.(2011)Povey, Ghoshal, Boulianne, Burget, Glembek, Goel,
  Hannemann, Motlicek, Qian, Schwarz et~al.}]{povey2011kaldi}
Daniel Povey, Arnab Ghoshal, Gilles Boulianne, Lukas Burget, Ondrej Glembek,
  Nagendra Goel, Mirko Hannemann, Petr Motlicek, Yanmin Qian, Petr Schwarz,
  et~al. 2011.
\newblock The {Kaldi} speech recognition toolkit.
\newblock In \emph{Proc. of the IEEE Workshop on Automatic Speech Recognition
  and Understanding}.

\bibitem[{Povey et~al.(2016)Povey, Peddinti, Galvez, Ghahremani, Manohar, Na,
  Wang, and Khudanpur}]{DBLP:conf/interspeech/PoveyPGGMNWK16}
Daniel Povey, Vijayaditya Peddinti, Daniel Galvez, Pegah Ghahremani, Vimal
  Manohar, Xingyu Na, Yiming Wang, and Sanjeev Khudanpur. 2016.
\newblock \href {https://doi.org/10.21437/Interspeech.2016-595} {Purely
  sequence-trained neural networks for {ASR} based on lattice-free {MMI}}.
\newblock In \emph{Proc. of the 17th Annual Conference of the International
  Speech Communication Association (INTERSPEECH), San Francisco, CA, USA,
  September 8-12, 2016}, pages 2751--2755.

\bibitem[{Sainath et~al.(2018)Sainath, Prabhavalkar, Kumar, Lee, Kannan,
  Rybach, Schogol, Nguyen, Li, Wu, Chen, and
  Chiu}]{DBLP:conf/icassp/SainathPKLKRSNL18}
Tara~N. Sainath, Rohit Prabhavalkar, Shankar Kumar, Seungji Lee, Anjuli Kannan,
  David Rybach, Vlad Schogol, Patrick Nguyen, Bo~Li, Yonghui Wu, Zhifeng Chen,
  and Chung{-}Cheng Chiu. 2018.
\newblock \href {https://doi.org/10.1109/ICASSP.2018.8462380} {No need for a
  lexicon? {Evaluating} the value of the pronunciation lexica in end-to-end
  models}.
\newblock In \emph{Proc. of the {IEEE} International Conference on Acoustics,
  Speech and Signal Processing ({ICASSP}), Calgary, AB, Canada, April 15-20,
  2018}, pages 5859--5863.

\bibitem[{Sennrich et~al.(2016)Sennrich, Haddow, and
  Birch}]{DBLP:conf/acl/SennrichHB16a}
Rico Sennrich, Barry Haddow, and Alexandra Birch. 2016.
\newblock \href {https://doi.org/10.18653/v1/p16-1162} {Neural machine
  translation of rare words with subword units}.
\newblock In \emph{Proc. of the 54th Annual Meeting of the Association for
  Computational Linguistics (ACL), August 7-12, 2016, Berlin, Germany}.

\bibitem[{Shi et~al.(2017)Shi, Hamdullah, Tang, Wang, and
  Zheng}]{DBLP:conf/apsipa/ShiHTWZ17}
Ying Shi, Askar Hamdullah, Zhiyuan Tang, Dong Wang, and Thomas~Fang Zheng.
  2017.
\newblock \href {https://doi.org/10.1109/APSIPA.2017.8282133} {A free {Kazakh}
  speech database and a speech recognition baseline}.
\newblock In \emph{Proc. of the Asia-Pacific Signal and Information Processing
  Association Annual Summit and Conference (APSIPA), Kuala Lumpur, Malaysia,
  December 12-15, 2017}, pages 745--748.

\bibitem[{Simonyan and Zisserman(2015)}]{DBLP:journals/corr/SimonyanZ14a}
Karen Simonyan and Andrew Zisserman. 2015.
\newblock \href {http://arxiv.org/abs/1409.1556} {Very deep convolutional
  networks for large-scale image recognition}.
\newblock In \emph{Proc. of the 3rd International Conference on Learning
  Representations (ICLR), San Diego, CA, USA, May 7-9, 2015}.

\bibitem[{Snow et~al.(2008)Snow, O'Connor, Jurafsky, and
  Ng}]{DBLP:conf/emnlp/SnowOJN08}
Rion Snow, Brendan O'Connor, Daniel Jurafsky, and Andrew~Y. Ng. 2008.
\newblock \href {https://www.aclweb.org/anthology/D08-1027/} {Cheap and fast -
  {But} is it good? {Evaluating} non-expert annotations for natural language
  tasks}.
\newblock In \emph{Proc. of the Conference on Empirical Methods in Natural
  Language Processing, (EMNLP), 25-27 October 2008, Honolulu, Hawaii, USA},
  pages 254--263.

\bibitem[{Stolcke(2002)}]{DBLP:conf/interspeech/Stolcke02}
Andreas Stolcke. 2002.
\newblock \href {http://www.isca-speech.org/archive/icslp\_2002/i02\_0901.html}
  {{SRILM} - {An} extensible language modeling toolkit}.
\newblock In \emph{Proc. of the International Conference on Spoken Language
  Processing (ICSLP), Denver, Colorado, USA, September 16-20, 2002}.

\bibitem[{Takamichi and Saruwatari(2018)}]{DBLP:conf/lrec/TakamichiS18}
Shinnosuke Takamichi and Hiroshi Saruwatari. 2018.
\newblock \href
  {http://www.lrec-conf.org/proceedings/lrec2018/summaries/67.html} {{CPJD}
  corpus: Crowdsourced parallel speech corpus of {Japanese} dialects}.
\newblock In \emph{Proc. of the 11th International Conference on Language
  Resources and Evaluation (LREC), Miyazaki, Japan, May 7-12, 2018}.

\bibitem[{Vaswani et~al.(2017)Vaswani, Shazeer, Parmar, Uszkoreit, Jones,
  Gomez, Kaiser, and Polosukhin}]{DBLP:conf/nips/VaswaniSPUJGKP17}
Ashish Vaswani, Noam Shazeer, Niki Parmar, Jakob Uszkoreit, Llion Jones,
  Aidan~N. Gomez, Lukasz Kaiser, and Illia Polosukhin. 2017.
\newblock \href {http://papers.nips.cc/paper/7181-attention-is-all-you-need}
  {Attention is all you need}.
\newblock In \emph{Proc. of the Annual Conference on Advances in Neural
  Information Processing Systems (NIPS), 4-9 December 2017, Long Beach, CA,
  {USA}}, pages 5998--6008.

\bibitem[{Watanabe et~al.(2018)Watanabe, Hori, Karita, Hayashi, Nishitoba,
  Unno, {Enrique Yalta Soplin}, Heymann, Wiesner, Chen, Renduchintala, and
  Ochiai}]{watanabe2018espnet}
Shinji Watanabe, Takaaki Hori, Shigeki Karita, Tomoki Hayashi, Jiro Nishitoba,
  Yuya Unno, Nelson {Enrique Yalta Soplin}, Jahn Heymann, Matthew Wiesner,
  Nanxin Chen, Adithya Renduchintala, and Tsubasa Ochiai. 2018.
\newblock \href {https://doi.org/10.21437/Interspeech.2018-1456} {{ESPnet}:
  {End}-to-end speech processing toolkit}.
\newblock In \emph{Proc. of the 19th Annual Conference of the International
  Speech Communication Association (INTERSPEECH), Hyderabad, India, 2-6
  September 2018}, pages 2207--2211.

\bibitem[{Yu and Deng(2014)}]{yu2016automatic}
Dong Yu and Li~Deng. 2014.
\newblock \emph{Automatic Speech Recognition: A Deep Learning Approach}.
\newblock Springer Publishing Company.

\bibitem[{Zhou et~al.(2020)Zhou, Michel, Irie, Kitza, Schl{\"{u}}ter, and
  Ney}]{DBLP:conf/icassp/ZhouMIKSN20}
Wei Zhou, Wilfried Michel, Kazuki Irie, Markus Kitza, Ralf Schl{\"{u}}ter, and
  Hermann Ney. 2020.
\newblock \href {https://doi.org/10.1109/ICASSP40776.2020.9053573} {The rwth
  asr system for ted-lium release 2: Improving hybrid hmm with specaugment}.
\newblock In \emph{2020 {IEEE} International Conference on Acoustics, Speech
  and Signal Processing, {ICASSP} 2020, Barcelona, Spain, May 4-8, 2020}, pages
  7839--7843. {IEEE}.

\end{thebibliography}
\bibliographystyle{acl_natbib}

\end{document}